\documentclass[useAMS,usenatbib]{mn2e}
\usepackage{amsmath}                                                            
\usepackage{amsfonts}
\usepackage{graphicx}
\usepackage{float}
\usepackage{morefloats}
\usepackage{listings}
\usepackage{color,hyperref}
\definecolor{linkcolor}{rgb}{0,0,0.25}
\hypersetup{
  colorlinks=true,        
  linkcolor=linkcolor,    
  citecolor=linkcolor,    
  filecolor=linkcolor,    
  urlcolor=linkcolor      
}

\title[Fast Galactic rotators in TGAS]
{Stars with fast Galactic rotation observed in $Gaia$ TGAS: a signature driven by the Perseus arm?}
\author[J. A. S. Hunt et al.]
  {\parbox{\textwidth}{Jason A. S. Hunt$^{1,2}$\thanks{E-mail: jason.hunt@dunlap.utoronto.ca}, Daisuke Kawata$^{1}$, Giacomo Monari$^{3}$, Robert J. J. Grand$^{4,5}$ Benoit Famaey$^{3}$ and Arnaud Siebert$^{3}$.}\vspace{0.5cm}
\\
$^{1}$ Mullard Space Science Laboratory, University College London,
Holmbury St. Mary, Dorking, Surrey, RH5 6NT, UK \\
$^{2}$ Dunlap Institute for Astronomy and Astrophysics, University of Toronto, Ontario, M5S 3H4, Canada \\
$^3$ Observatoire astronomique de Strasbourg, Universit\'e de Strasbourg, CNRS UMR 7550, 11 rue de l'Universit\'e, 67000 Strasbourg, France\\
$^4$ Heidelberger Institut f\"ur Theoretische Studien, Schloss-Wolfsbrunnenweg 35, 69118 Heidelberg, Germany\\
$^5$ Zentrum f\"ur Astronomie der Universit\"at Heidelberg, Astronomisches Recheninstitut, M\"onchhofstr. 12-14, 69120 Heidelberg, Germany
}
\date{Accepted 2016 December 19. Received 2016 December 16; in original form 2016 November 2} 

\pagerange{\pageref{firstpage}--\pageref{lastpage}}
\pubyear{2016}

\begin{document}

\maketitle

\label{firstpage}

\begin{abstract}
We report on the detection of a small overdensity of stars in velocity space with systematically higher Galactocentric rotation velocity than the Sun by about 20~km~s$^{-1}$ in the $Gaia$ Data Release 1 Tycho-Gaia astrometric solution (TGAS) data. We find these fast Galactic rotators more clearly outside of the Solar radius, compared to inside of the Solar radius. In addition, the velocity of the fast Galactic rotators is independent of the Galactocentric distance up to $R-R_{\odot}\sim0.6$ kpc. Comparing with numerical models, we qualitatively discuss that a possible cause of this feature is the co-rotation resonance of the Perseus spiral arm, where the stars in peri-centre phase in the trailing side of the Perseus spiral arm experience an extended period of acceleration owing to the torque from the Perseus arm. 
\end{abstract}

\begin{keywords}
methods: $N$-body simulations --- methods: numerical --- galaxies: structure
--- galaxies: kinematics and dynamics --- The Galaxy: structure
\end{keywords}

\section{Introduction}
\label{intro-sec}

Milky way astronomy is currently entering an exciting era with the recent first data release  \citep[{\it Gaia} DR1,][]{GaiaDR1} from the European Space Agency (ESA)'s $Gaia$ mission \citep{GaiaMission}. $Gaia$ DR1 contains the Tycho-$Gaia$ Astrometric Solution \citep[TGAS,][]{Michalik+15,Lindegren16+} which provides positions, parallaxes and proper motions ($\alpha$, $\delta$, $\pi$, $\mu_{\alpha}$, $\mu_{\delta}$) for around 2 million stars using data from the $Tycho$-2
catalogue \citep[e.g.][]{P97,HFMUCetal00} to provide a baseline of approximately 30 years
upon which to calculate astrometric values for stars in common between $Tycho$-2 and $Gaia$. This enables us to explore local dynamics in unprecedented detail \citep[e.g.][]{Bovy16}. Additionally, to provide the full 6 dimensional phase space measurements, TGAS can be cross-matched with ground based spectroscopic surveys such as the Radial Velocity Experiment \citep[RAVE,][]{Sea06} in the Southern hemisphere and/or the Large Sky Area Multi-Object Fibre Spectroscopic Telescope \citep[LAMOST,][]{ZZCJD12} and the Apache Point Observatory Galactic Evolution Experiment \citep[APOGEE,][]{MAPOGEE16} in the Northern hemisphere to gain radial velocity measurements for stars in both catalogues. This has already been demonstrated to grant new insights into stellar dynamics \citep{APKC16,Hunt+16,Monari+16}.

In this {\it Letter}, we present a first detection of a small group of stars in the TGAS data whose Galactocentric rotation velocity is systematically higher than the Sun, independent of radius. Hereafter we call these stars fast Galactic rotators which appear clearly only outside the Solar radius. In Section \ref{TGAS} we describe our treatment of the data and the observed feature. In Section \ref{sim} we compare the data with numerical models and qualitatively discuss that the fast Galactic rotators found in the TGAS data could be interpreted as a signature of the co-rotation resonance of the Perseus arm. In Section \ref{Summary} we summarise the results.

\section{Fast Galactic Rotators in the TGAS data}
\label{TGAS}


We analyse the rotation velocity of stars from the Gaia DR1 TGAS data. The TGAS catalogue provides parallaxes, $\pi$, and proper motions ($\mu_{\alpha}$, $\mu_{\delta}$), but does not include any radial velocity information which would allow us to calculate Galactocentric rotation velocity, $v_{\text{rot}}$. Such radial velocities can be obtained for a subset of stars by combining TGAS with spectroscopic surveys \citep[e.g.,][]{Hunt+16,Monari+16}, but here we choose to analyse TGAS data on their own. We can for instance examine the velocity in the direction of Galactic longitude, $v_l$, which at $(l,b)=(180^\circ,0)$ and $(0,0)$, provides comparable information to $v_{\text{rot}}$, with respect to the Solar motion. We define $v_l=4.74047\mu_{l^*}/\pi$, where $\mu_{l^*}=\mu_l\cos(b)$ is the proper motion for the longitude direction in true arc.


\begin{figure}
\centering
\includegraphics[width=\hsize]{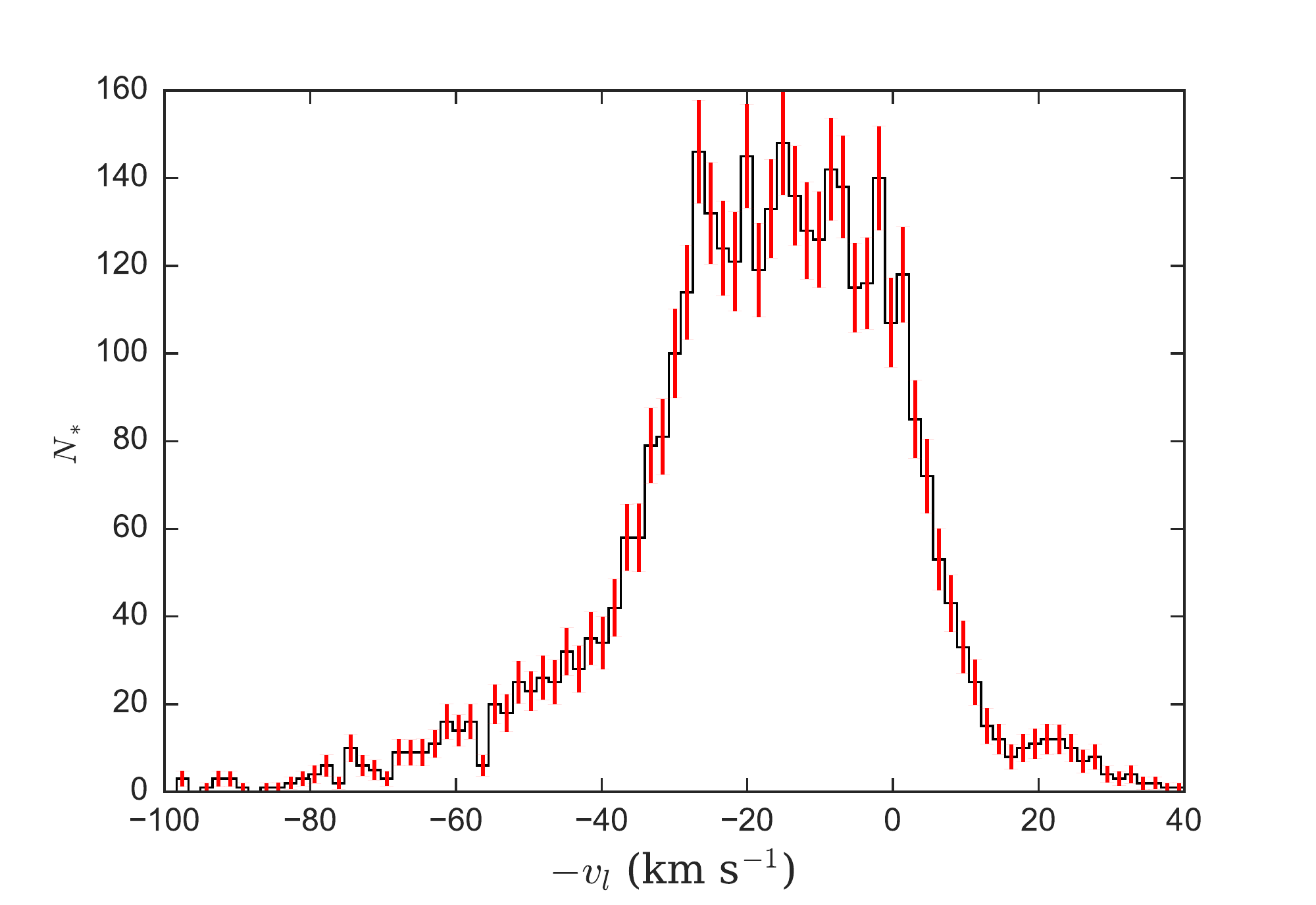}
\caption{Histogram of $-v_l$ (km s$^{-1}$) for stars within the line-of-sight area of $(l,b)=(180^\circ,0)\pm(10^\circ,5^\circ)$ and with $\sigma_{\pi}/\pi\leq0.15$.}
\label{180Hist}
\end{figure}

\begin{figure}
\centering
\includegraphics[width=\hsize]{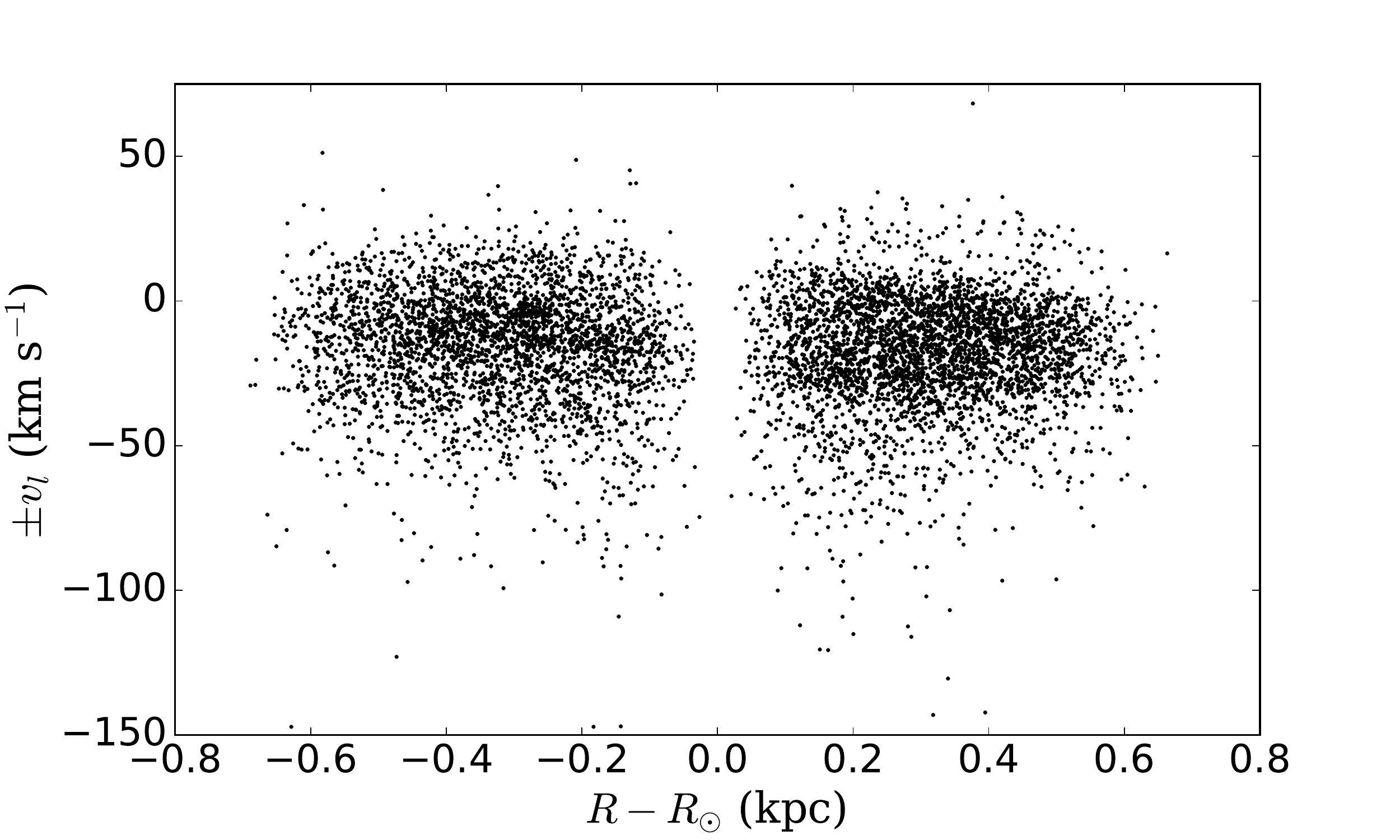}
\caption{$-v_l$ (km s$^{-1}$) against $R-R_0$ for stars within the line-of-sight area of $(l,b)=(180^\circ,0)\pm(10^\circ,5^\circ)$ (right hand grouping) and $+v_l$ for stars within the area of $(l,b)=(0,0)\pm(10^\circ,5^\circ)$ (left hand grouping).}
\label{scatter}
\end{figure}

Fig. \ref{180Hist} shows the histogram of $-v_{l}$ (km s$^{-1}$) for stars within the line-of-sight area of $(l,b)=(180^\circ,0)\pm(10^\circ,5^\circ)$. In this {\it Letter}, we limit our sample to only stars whose fractional parallax error is less than 15 per cent, i.e. $\sigma_{\pi}/\pi\leq0.15$. Along this particular line-of-sight, $-v_l$ roughly corresponds to the rotation velocity with respect to the Sun, $v_{\text{rot}}$, and higher $-v_l$ indicates faster $v_{\text{rot}}$. The error bars are $1\sigma$ and were calculated using the bootstrapping technique, taking 1000 samples with replacement. There is a small, but clear `bump' around $-v_l=20$ km s$^{-1}$, which corresponds to a group of stars with $-v_l$ about 20 km~s$^{-1}$ faster than the Sun (hereafter we call this group of stars `fast Galactic rotators'). Note that the current observational data suggest that the Sun rotates faster than the circular velocity at the Solar radius by about $V_{\odot}=12.24$ km s$^{-1}$ \citep{SBD10}.  In addition, the mean rotation of the stars is slower than the circular velocity  by the asymmetric drift. Therefore, these fast Galactic rotators are rotating significantly faster than the mean rotation of the stars.



Fig. \ref{scatter} shows $-v_l$ (km s$^{-1}$) against $R-R_0$ for stars within the line-of-sight area of $(l,b)=(180^\circ,0)\pm(10^\circ,5^\circ)$ (right hand grouping) and $v_l$ for stars within $(l,b)=(0,0)\pm(10^\circ,5^\circ)$ (left hand grouping), i.e. positive velocity is in the direction of Galactic rotation for both samples. Here, $R$ is the Galactocentric radius of stars and $R_0$ indicates the Solar radius. 
These fast Galactic rotators, with $-v_l\sim20$ km~s$^{-1}$, are clearly visible above the bulk of the sample in the right hand grouping, i.e. outside of the Solar radius. On the other hand, the fast Galactic rotators are missing from the left hand group at $(l,b)=(0,0)\pm(10^\circ,5^\circ)$, i.e. inside of the Solar radius. 

This is to our knowledge, the first discovery of these fast Galactic rotators just outside of the Solar radius, and the signature extends at least from $R-R_0=0.1$ to 0.6 kpc. We checked that the parallax and proper motion errors of the fast Galactic rotators are not systematically high compared to the other stars in the sample. We also checked that the `excess astrometric noise' is not high for these stars. In other words, they are unlikely to be binary stars \citep[see][]{Lindegren16+}. In the next section, we will present one possible cause of this feature, namely that this could be an indication of the co-rotation resonance of the Perseus spiral arm. 



\section{Signature driven by the Perseus spiral arm?}
\label{sim}

The kinematics of the Milky Way, along with other spiral galaxies, is likely heavily influenced by the spiral arms themselves. In this section, we qualitatively compare the data with numerical models, and propose that the fast Galactic rotators found in the TGAS data can be driven by the Perseus spiral arm, as one possible mechanism. 

\cite{LS64} proposed a solution to the so-called `winding dilemma' by treating the spiral structure as density wave features that rotate rigidly with a pattern speed that is constant with radius, irrespective of the rotation velocity of the stars themselves. However, $N$-body simulations have as of yet been unable to reproduce spiral arms as long-lived single modes \citep[e.g.][]{S11,DB14}, instead showing spiral arm features which are short-lived but recurrent with pattern speeds that match the rotation of the stars, i.e. co-rotating at all radii \citep[e.g.][]{SC84,GKC12,RFetal13}. 
One of the interpretations of the $N$-body simulations is that while the spiral arm features themselves are transient, their evolution in configuration space may be driven by several spiral modes present in the disc that are in fact longer-lived standing wave oscillations (lasting for about 10 rotations) similar to the original density wave theory \citep[e.g.][]{MFQDCVEB12,RDL13,SC14}. 

 We therefore consider two types of the numerical models, an $N$-body model from \cite{KHGPC14} and a test particle model with rigidly-rotating spiral arms from \cite{MFSGKB16}, and analyse the rotation velocity distribution in the direction of $(l,b)=(180^\circ,0)$. Note that these models are both also analysed in \cite{GBKHFSMC15}.



The details of the numerical simulation code, and the galaxy model of the N-body model are described in \cite{KHGPC14}. The galaxy is set up in isolated conditions, and consists of a gas and stellar disc but no bulge component. 
The model shows transient spiral arm features, co-rotating with the stars at all radii as discussed in \cite{KHGPC14}. We examine a snapshot at an earlier time from this model with a strong spiral arm as shown in \cite{KHGPC14} (K14a) and a snapshot at a later time from this model with a weak spiral arm as shown in \cite{GBKHFSMC15} (K14b). A comparison of this galaxy to the Milky Way including measurement of its age/velocity dispersion, bar strength and the pitch angle of the spiral arms are given for K14a in \cite{HKGMPC15} and for K14b in \cite{GBKHFSMC15}. We assume a solar radius of $R_0=8$~kpc for both K14a and K14b.

The test particle model we use is model S2 from \cite{MFSGKB16}, henceforth M16.
The details of the model are given in \cite{MFSGKB16}. The gravitational potential used is comprised of an axisymmetric part \citep[Model 1 from][]{GalD08}, along with a rigidly-rotating two armed spiral perturbation. The axisymmetric component consists of two spherical components, a dark halo, a bulge and three disc components; thin, thick and interstellar medium (ISM). 
The spiral perturbation is of the form given in \cite{CG02}, with the pattern speed and pitch angle derived by \cite{Siebert+12}. The co-rotation radius of the spiral arm is 11.49~kpc. We assumed the Solar radius of $R_0=8$~kpc.


Although we wish to compare $v_l$ directly between the TGAS data and our models, our models are too low resolution to analyse the velocity distribution in such a small area which we explore in Section~\ref{TGAS}. Thus, we expand the selection of the model particles to cover a 1 kpc cube such that $0<R-R_{\odot}<1$ kpc and $-0.5<y,z<0.5$ kpc, where the $y$-direction is the direction of rotation and the $z$-direction points toward the North Galactic Pole. For the models, we have 6 dimensional position and velocity information, and therefore we do not need to use $-v_l$ as a proxy for the rotation velocity with respect to the Sun, $v_{\text{rot}}$, as we did for the TGAS data. Hence, we directly analysed the rotation velocity of the $i$-th particle with respect to the observer as $v_{\text{rot},i}=v_{\phi,i}-v_{\text{circ}}-V_{\odot}$, where  $v_{\phi,i}$ is the Galactocentric rotation velocity of the $i$-th particle, $v_{\text{circ}}$ is the circular velocity of the model disc at $R_0$ and $V_{\odot}=12.24$ km s$^{-1}$ \citep{SBD10} is the solar proper motion assumed. 


\begin{figure*}
\centering
\includegraphics[width=\hsize]{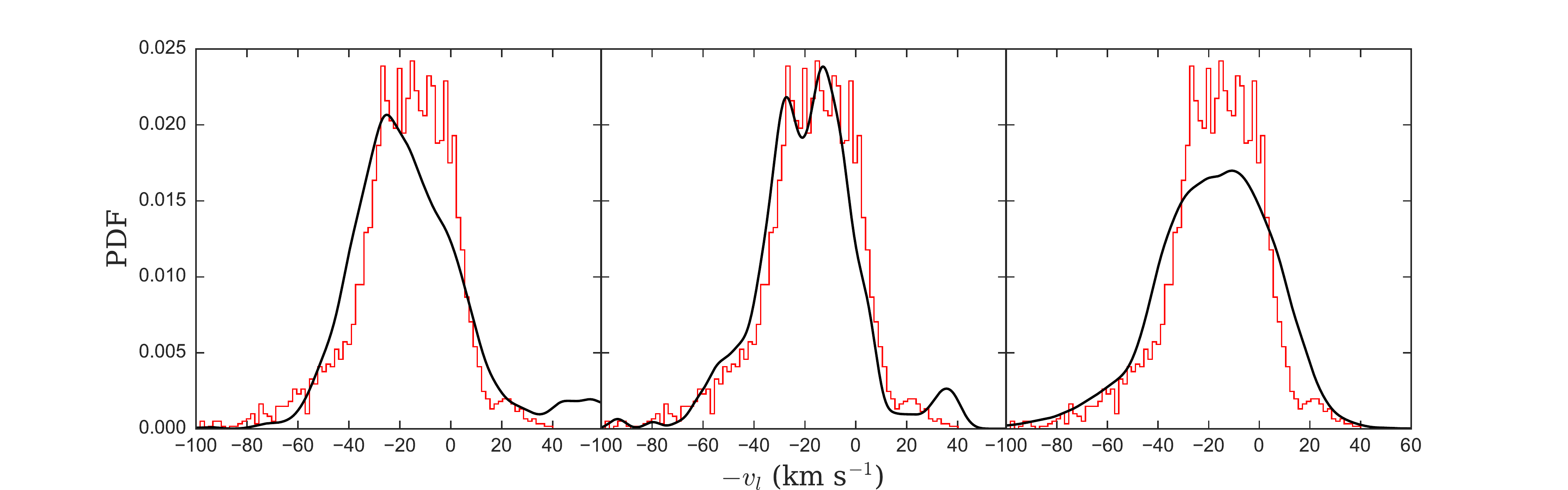}
\caption{Probability Distribution Function (PDF) for $v_{\text{rot}}$ (km s$^{-1}$) for model particles in K14a (left), K14b (centre) and M16 (right) for the line of sight $(l,b)=(180,0)$ on the near trailing side of the spiral arm. Overlaid on a histogram of $v_l$ (km s$^{-1}$) constructed from the TGAS data. $v_{\text{rot}}$ and $v_l$ are roughly equivalent at $(l,b)=(180,0)$.}
\label{vKDE}
\end{figure*}

Fig. \ref{vKDE} shows the Probability Distribution Function (PDF) for $v_{\mathrm{rot}}$ for model particles in K14a (left), K14b (centre) and M16 (right) for the line of sight $(l,b)=(180^\circ,0)$, overlaid on a histogram of $-v_l$ of the TGAS data as shown in Fig.~\ref{180Hist}. The line shows the distribution of $v_{\mathrm{rot}}$ of the model particles in the volume described above. The left and central panels of Fig. \ref{vKDE} show clearly a small secondary high velocity peak in the distribution at approximately 50 km s$^{-1}$ and 25 km s$^{-1}$ respectively. In contrast, the PDF of M16 in the right panel shows little structure in the distribution of $v_{\mathrm{rot}}$. There is no evidence of a separate peak for the high velocity population. Instead it exhibits a more symmetric spread of the rotation velocity, which clearly contradicts the observed distribution.

The strong arm model K14a in the left panel is also clearly inconsistent with the data, because K14a shows fast Galactic rotators with $v_l\approx50$ km s$^{-1}$ which is much faster than the observed fast Galactic rotators. Model K14b in the central panel also overestimates the velocity of the fast Galactic rotators. However, the comparison between Models K14a and K14b suggests that the velocity of the `bump' of the fast Galactic rotators gets slower with a weaker arm. Therefore, it is likely an indication that the arm in Model K14b is still too strong compared to the spiral arms of the Milky Way, and it is reasonable to assume that a weaker arm would reproduce the data nicely. 
It is pleasing to note that apart from a slight offset of the location of the fast Galactic rotators, Model K14b is in good agreement to the data, including the slight double peak in the main distribution and a sharper cut off in the faster rotation side compared to the tail in the slower rotation side. 



In \cite{KHGPC14} we showed that these fast Galactic rotators observed in the $N$-body models are caused by stars which are initially close to apo-centre phase on the leading side of the spiral arm and are decelerated by the arm. After the spiral arm has overtaken these stars, they are accelerated by the spiral arm while approaching their peri-centre phase, which leads to a group of fast Galactic rotators on the trailing side of the spiral arm. This is because the spiral arms seen in $N$-body simulations are co-rotating, and thus the stars can stay close to the spiral arm for an extended period of time. This can cause them to be significantly accelerated on the trailing side of the spiral arm\footnote{Note that as demonstrated in \citet{GKC12} the non-linear growth of the spiral arm is resulting from a mechanism similar to swing-amplification \citep{JT66}, but which differs in terms of the evolution, and occurs at a wide range of radii.}. In other words, this feature is driven by the co-rotation resonance of the spiral arm. In M16, we do not find any feature, because the co-rotation radius ($R_c=11.49$~kpc) is far away from the assumed Solar radius.

At $l=180^\circ$ the Perseus arm is located approximately 2 kpc outwards from the Sun \citep[e.g.][]{Retal09}. Thus, we are observing the trailing side of the Perseus arm. Therefore, we speculate that the fast Galactic rotators found in the TGAS data could be the stars driven by the co-rotation resonance of the Perseus arm in the way described above. On the other hand, at $l=0$ the Perseus arm is further away, and therefore the influence of the arm is not strong enough to induce the fast Galactic rotators. If so, the Perseus arm could be co-rotating and transient as observed in $N$-body simulations. However, this feature could perhaps be equally well explained by classic spiral density wave theory where the spiral arm features are rigidly rotating and there is a specific co-rotation radius. For example, if there is a density wave with co-rotation radius just outside the Solar radius, we would also expect to observe these fast Galactic rotators outside the Solar radius. However, the overdensity in velocity space would move with radius at a given azimuth.

\begin{figure*}
\centering
\includegraphics[width=\hsize]{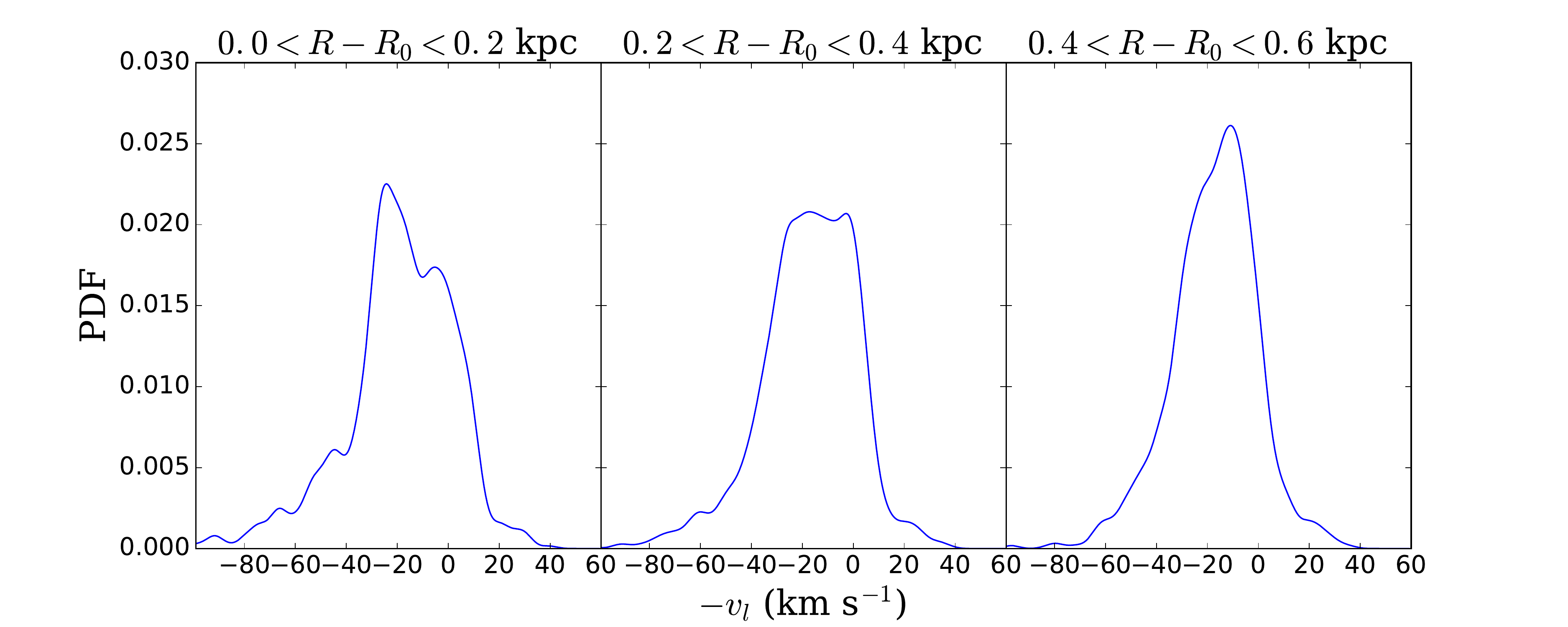}
\caption{Probability Distribution Function (PDF) for $-v_{l}$ (km s$^{-1}$) for $0.0<R-R_0<0.2$ kpc (left), $0.2<R-R_0<0.4$ kpc (centre) and $0.4<R-R_0<0.6$ kpc (right).}
\label{vldist}
\end{figure*}

To distinguish these theories of the dynamics of spiral structure, we need to measure the rotation velocity distribution at different radii of the Perseus arm itself, as discussed in \cite{KHGPC14} and \cite{HKGMPC15}. If the Perseus arm is like co-rotating spiral arms seen in $N$-body simulations, we expect that the feature found in this {\it Letter} will be observed on the trailing side of the Perseus arm at different radii, i.e. different $l$, because the spiral arm features are co-rotating at all radii. Unfortunately, the volume where TGAS provides accurate enough parallax and proper motion is too small for us to analyse the rotation velocity distribution at different $l$. 


However, we can measure the `bump' feature of the fast Galactic rotators with varying radius up to approximately 0.6 kpc. Fig. \ref{vldist} shows the distribution of $-v_l$ (km s$^{-1}$) for $0.0<R-R_0<0.2$ kpc (left), $0.2<R-R_0<0.4$ kpc (centre) and $0.4<R-R_0<0.6$ kpc (right). Although the variation in the structure of the velocity distribution is clear, the position of the bump of fast Galactic rotators around $-v_l=20$ km~s$^{-1}$ does not alter velocity with increasing distance. The constant rotation speed of the feature for $0.0\leq R-R_0\leq 0.5$ kpc can also be seen in Fig.~3 of \citet{Monari+16} at $v_{\rm \phi}\sim275$ km~s$^{-1}$. They assume a Solar Galactocentric rotation speed of 254.6 km~s$^{-1}$ and hence this is consistent with the 20 km~s$^{-1}$ fast Galactic rotators observed in this {\it Letter}. The combination of the non-variation of the feature with distance, and the fact that it is also observed in the TGAS+LAMOST data shown in \cite{Monari+16} make it unlikely to be a feature of noise alone.

If this feature were caused by a single resonance, e.g. relating to the bar, it would be expected to vary with distance in a fashion similar to the $Hercules$ stream, as shown in \cite{Monari+16}. Note that we also observe the $Hercules$ stream in Fig. \ref{scatter} as an overdensity defined by a gap going from $-v_l\sim -30$ km s$^{-1}$ at $R-R_0 \sim 0.1$ kpc to $-v_l\sim -45$ km s$^{-1}$ at $R-R_0 \sim 0.3$ kpc, and then fading away, although it is not clearly visible in Figure 1 because the gap varies with radius.


\section{Summary}
\label{Summary}

Analysing the newly released $Gaia$ TGAS data, we found a group of stars which have systematically high rotation velocity just outside of the Solar radius. We compare them with snapshots of an $N$-body simulation and a test-particle simulation with spiral arms, and find that these fast Galactic rotators can be explained by peri-centre phase stars being accelerated by the Perseus arm for an extended period, owing to the co-rotation resonance. This supports the scenarios of the Perseus arm being either a transient and co-rotating spiral arm like seen in $N$-body simulations or a density wave whose co-rotation radius is just outside of the Solar radius. From the current data, we cannot distinguish between these two possibilities, although the non-variation of the feature in velocity space with respect to radius does not argue in favour of a single co-rotation radius.

There are also other possible explanations for this feature not necessarily linked to non-axisymmetric patterns such as spiral arms, for example, these fast Galactic rotators could also be made of high eccentricity stars triggered by an interaction event in the outer disc, whose peri-centre of their orbits is close to the Solar radius, hence at high rotational velocity. Although in this case, the location of the bump in $v_l$ would be expected to vary with distance, as it would if it is a resonance feature of the bar or of a single static spiral density wave. The fact that the feature in $v_l$ remains constant with distance makes either of these explanations unlikely.

$Gaia$ DR2 will allow us to examine stellar dynamics over a much larger area of the Galaxy, and with far greater accuracy. We will be able to probe different lines-of-sight using $Gaia$ radial velocities, and also observe the far leading side of the Perseus arm. This information will provide a crucial test for these competing spiral arm models as demonstrated in \cite{HKGMPC15}.

\section*{Acknowledgements}
We thank James Binney for his useful suggestions and  discussion.
This study was developed during the Gaia Challenge IV workshop hosted and supported by  Nordic  Institute for Theoretical Physics.
JH is supported by a Dunlap Fellowship at the Dunlap Institute for Astronomy \& Astrophysics, funded through an endowment established by the Dunlap family and the University of Toronto. 
DK acknowledges the support of the UK's Science \& Technology 
Facilities Council (STFC Grant ST/K000977/1 and ST/N000811/1).  
Some of the numerical Galaxy models for this paper were simulated on the UCL facility Grace, and the DiRAC Facilities (through the COSMOS and MSSL-Astro consortium) jointly 
funded by STFC and the Large Facilities Capital Fund of BIS.  We also acknowledge 
PRACE for awarding us access to their Tier-1 facilities.
This work has made use of data from the European Space Agency (ESA)
mission $Gaia$ (\url{http://www.cosmos.esa.int/gaia}), processed
by the $Gaia$ Data Processing and Analysis Consortium (DPAC,
\url{http://www.cosmos.esa.int/web/gaia/dpac/consortium}). Funding for the DPAC has been provided by national institutions, in particular the institutions participating in the $Gaia$ Multilateral
Agreement.

\bibliographystyle{mn2e}
\bibliography{ref2}

\label{lastpage}
\end{document}